\documentclass[a4paper]{article}

\usepackage{INTERSPEECH2020}
\usepackage{color}

\newcommand{\AttName}{{\ModelName} attention}
\newcommand{\ModelName}{MoBoAligner}
\newcommand{\NarName}{{FastSpeech}}
\newcommand{\TokenName}{text token}
\newcommand{\FA}{Forced alignment}

\newcommand{\TfMos}{$4.18$} 
\newcommand{\FsMos}{$3.44$}
\newcommand{\MbMos}{$3.74$}
\newcommand{\GtMos}{$4.48$}
\newcommand{\RcMos}{$4.61$}

\title{\ModelName:  a Neural Alignment Model for Non-autoregressive TTS with Monotonic Boundary Search}
\name{Naihan Li$^1$, Shujie Liu$^2$, Yanqing Liu$^3$, Sheng Zhao$^3$, Ming Liu$^1$ and Ming Zhou$^2$}
\address{
  $^1$University of Electronic Science and Technology of China\\
  $^2$ Microsoft Research Asia, Beijing, China
  $^3$Microsoft Speech and Language Group}
\email{lnhzsbs1994@163.com, shujliu@microsoft.com, yanqliu@microsoft.com, szhao@microsoft.com, yanqliu@microsoft.com, mingzhou@microsoft.com}
\email{lnhzsbs1994@163.com, \{shujliu, yanqliu, szhao, mingzhou\}@microsoft.com, csmliu@uestc.edu.cn}

\begin{document}

\maketitle
\begin{abstract}
To speed up the inference of neural speech synthesis, non-autoregressive models receive increasing attention recently. In non-autoregressive models, additional durations of text tokens are required to make a hard alignment between the encoder and the decoder. 
The duration-based alignment plays a crucial role since it controls the correspondence between text tokens and spectrum frames and determines the rhythm and speed of synthesized audio.
To get better duration-based alignment and improve the quality of non-autoregressive speech synthesis, in this paper, we propose a novel neural alignment model named {\ModelName}. Given the pairs of the text and mel spectrum, {\ModelName} tries to identify the boundaries of text tokens in the given mel spectrum frames based on the token-frame similarity in the neural semantic space with an end-to-end framework. With these boundaries, durations can be extracted and used in the training of non-autoregressive TTS models.
Compared with the duration extracted by TransformerTTS, {\ModelName} brings improvement for the non-autoregressive TTS model on MOS (our {\MbMos} comparing to baseline's {\FsMos}). Besides, {\ModelName} is task-specified and lightweight, which reduces the parameter number by 45\% and the training time consuming by 30\%.

\end{abstract}

\section{Introduction}\label{sec:intro}
Speech synthesis (text to speech, TTS) plays a pivotal role with a wide application in the speech-related scenario. After a step made from statistic and parametric TTS models \cite{tokuda2000speech,zen2009statistical,ze2013statistical,tokuda2013speech}, neural TTS models become the mainstay owing to the advance of deep learning.  There are two modules in the neural TTS pipeline, the acoustic model to convert the input text to spectrum, and the vocoder to synthesize the audio conditioning on the spectrum.
The vocoders include Wavenet \cite{oord2016wavenet}, Parallel Wavenet \cite{oord2017parallel}, WaveRNN \cite{kalchbrenner2018efficient}, WaveGlow \cite{prenger2019waveglow}, Parallel WaveGAN \cite{yamamoto2019parallel}, etc. 
As for the acoustic models, Tacotron2 \cite{shen2017natural} and TransformerTTS \cite{li2018transformertts} are two sequence-to-sequence based acoustic models that can synthesize mel spectrums in an autoregressive manner. The autoregressive dependencies between spectrum frames constrain parallel computing, resulting in low efficiency during inference. 
Therefore, a non-autoregressive architecture named FastSpeech \cite{ren2019fastspeech} is proposed.

\begin{figure}[tb]
  \centering
    \includegraphics[width=0.8\linewidth]{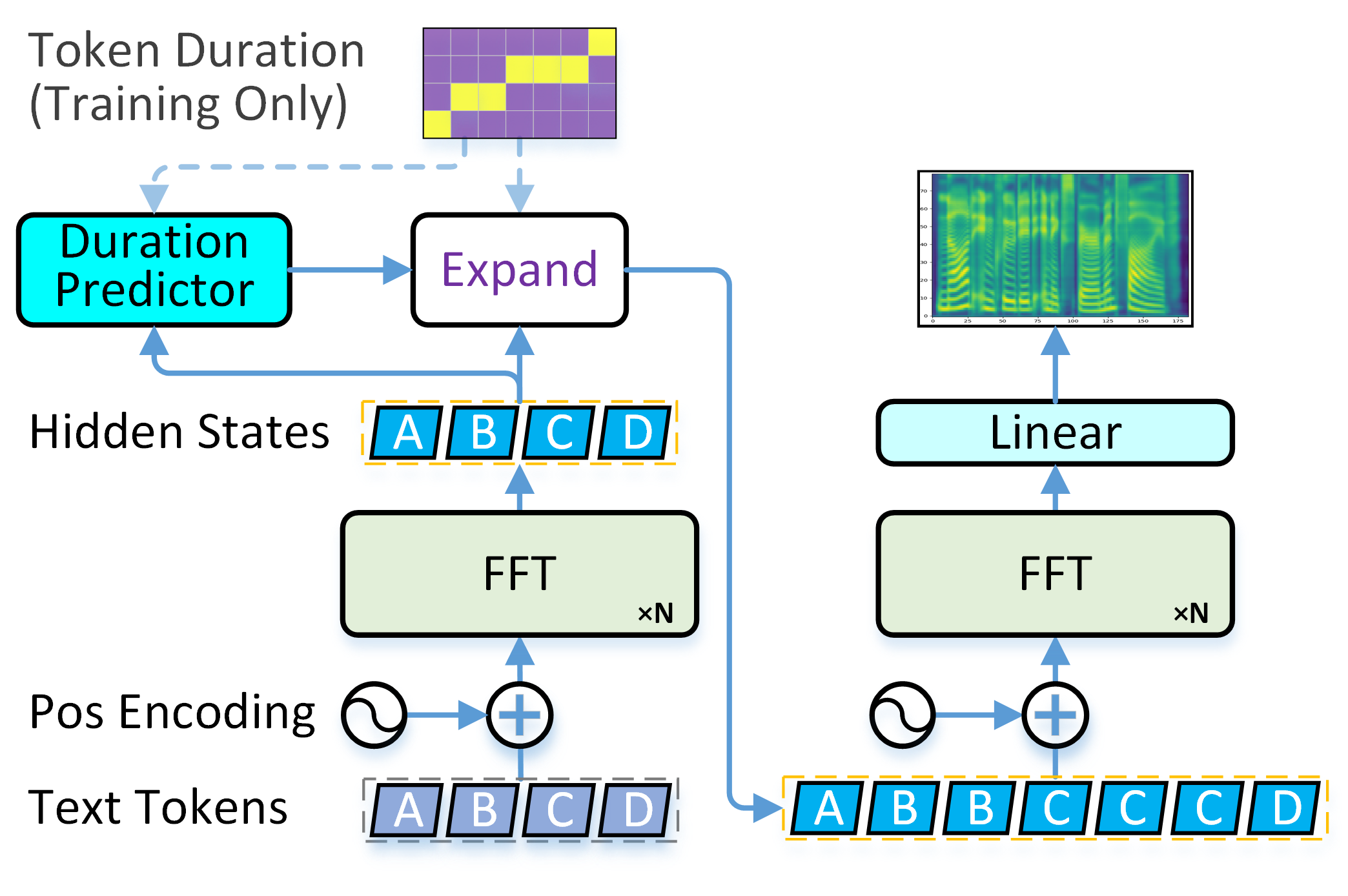}
  \caption{{\NarName}. Schematically, the input and the token duration is the same as it is in Fig. \ref{fig:model}, which is processed by the left FFT (Transformer encoder), expanded and fed into the right FFT (Transformer decoder) to generate the mel spectrum. The token duration is required only for training. }
  \label{fig:FFT}
\end{figure}

FastSpeech employs a novel Transformer-based model, which is called "Feed-forward Transformer", as shown in Fig. \ref{fig:FFT}. An encoder is firstly leveraged to project the input text into hidden states, which are expanded according to the token durations. After that, the expanded hidden states are consumed by the decoder to generate the mel spectrum. 
To expand the encoder hidden states and train the duration predictor in the training procedure of non-autoregressive TTS models, token durations are needed as shown in the top of Fig. \ref{fig:FFT}.
These durations play critical roles for two reasons: 1) they directly control the correspondence between text tokens and spectrum frames, thus the decoder learns to pronounce the given token and back-propagate the gradient to corresponding encoder hidden states during training; 2) accurate duration supervision for the duration predictor can benefit the rhythm and speed of the synthesized audio during inference.

{\cite{ren2019fastspeech}} leverages a well-trained autoregressive TransformerTTS to provide the duration of each text token by the encoder-decoder attention. Specifically, based on the similarity to a diagonal matrix, one alignment matrix is selected from all the alignments of the multi-head encoder-decoder attention. In this alignment, one frame is counted to a text token if this token has the largest share in this frame's attention weights. 
However, since the attention mechanism in TransformerTTS is not designed for duration prediction, using it as supervision to train the duration predictor brings two problems: 1) The attention weights of each frame are not concentrated on one single token; instead, they are dispersed to obtain a broad context of the input text. Although the text token being pronounced is allocated with the largest share in most cases, there is still some noise which can disturb the extracted duration. 
2) For the speech synthesis task, the alignment of the encoder and decoder hidden states should be monotonic, and the aligned frames of one token should be continuous. 
The alignment of TransformerTTS sometimes violates these two properties, causing poor duration accuracy.
With inaccurate duration as training data, the duration predictor may be biased and harm the rhythm of the synthesized audio, and the wrong encoder hidden states may be used to reconstruct the mel spectrum.

Following the properties of monotonicity and continuity, we are inspired by the monotonic alignment \cite{raffel2017online}, and propose {\ModelName}, a novel monotonic aligner for non-autoregressive TTS.
{\ModelName} leverages two separate encoders to project both the input text and the mel spectrum into the same semantic space, after that these two sequences are aligned with a novel attention mechanism named {\ModelName} attention. {\ModelName} attention can monotonically scan the mel spectrum frames according to the text tokens, and spot the token boundaries in these frames. According to the alignment, the text hidden states are expanded to frame-level and reconstruct the mel spectrum in an end-to-end framework, which is under the instruction of Mean Squared Error (MSE) loss. 
To evaluate the quality of extracted durations and evaluate our proposed \ModelName, we train non-autoregressive models with different duration sources and conduct mean opinion score (MOS) tests. 
Our contributions are summarized as follows:
\begin{enumerate}
    \item We propose {\ModelName}, a neural alignment model which can monotonically scan the mel spectrum and search the boundaries of the given text token sequence, and form the text-to-mel alignment based on these boundaries.
    \item The durations extracted by {\ModelName} are more accurate, which brings an improvement for the non-autoregressive TTS model on MOS ({our {\MbMos} comparing to baseline's {\FsMos}}).
    \item Comparing to TransformerTTS, our model is task-specified and lightweight, which reduces the parameter number by 45\% and reduces the training time consuming by 30\%.
\end{enumerate}

\begin{figure}[tb]
  \centering
    \includegraphics[width=0.60\linewidth]{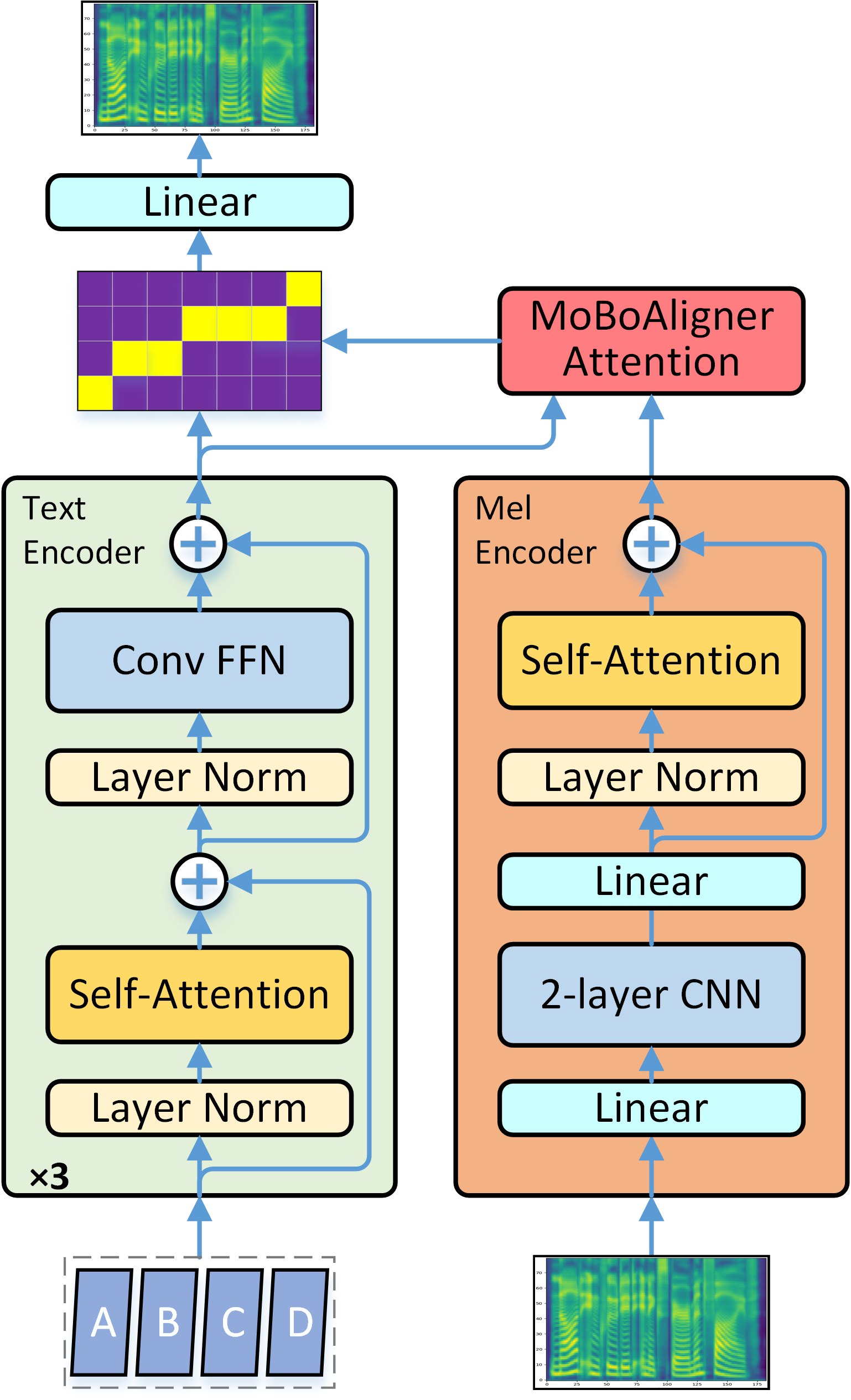}
  \caption{The architecture of \ModelName. Schematically, there are four tokens A, B, C and D in the input text, of which the durations are 1, 2, 3 and 1 as shown in the alignment.}
  \label{fig:model}
\end{figure}

\section{{\ModelName}}

As mentioned in the introduction, the non-autoregressive TTS model requires additional duration inputs, obtained by a TransformerTTS as designed in \cite{ren2019fastspeech}, which may not be  suitable for this task and have some disadvantages. 
To generate better alignment results for the training of duration predictor, in this section, we propose a novel neural aligner {\ModelName}, which can monotonically search the {\TokenName} boundaries in the mel spectrum, as shown in Fig. \ref{fig:model}. 
In {\ModelName}, the input text and mel spectrum are processed by the text encoder and mel encoder respectively. After that, {\AttName} (will be introduced in Section \ref{sec:\AttName}) is employed to monotonically align these two sequences and produce the alignment. Based on the alignment, the text hidden states are expanded to frame-level and reconstruct the mel spectrum by linear projection. Mean squared error (MSE) loss is employed as cost function, and the alignment between the text and the mel spectrum is automatically learned to minimize this loss. In the following of this section, we will introduce the text and mel encoders, and our {\AttName} will be introduced in Section \ref{sec:\AttName}

\subsection{Text encoder}
Each {\TokenName} is firstly embedded with a 512-dim vector, which is added with a scaled positional embedding is added, then fed into three Transformer FFT blocks. Here we change the fully-connected (FC) layers in the FFN with convolutional layers, of which the kernel size is 3. All attention sizes are 512, the head number is 8, and the hidden size of Conv FFN is 2048. 

\subsection{Mel encoder}
The mel spectrum is processed by a 2-layer CNN (channel number is 256, dilation is 2, kernel size is 3, followed by ReLU and dropout layers), which is designed for extracting the local context and strengthening the boundary information. A leading linear projection (unit size from 80 to 256, followed by the ReLU and dropout layers) and a tailing linear projection (unit size from 256 to 512) are used for dimension consistency, and a scaled positional embedding is added after the tailing linear projection. Finally, a self-attention layer is added to provide a sequence-level context.

\section{{\AttName}} \label{sec:\AttName}

\begin{figure}[tb]
  \centering
    \includegraphics[width=0.70\linewidth]{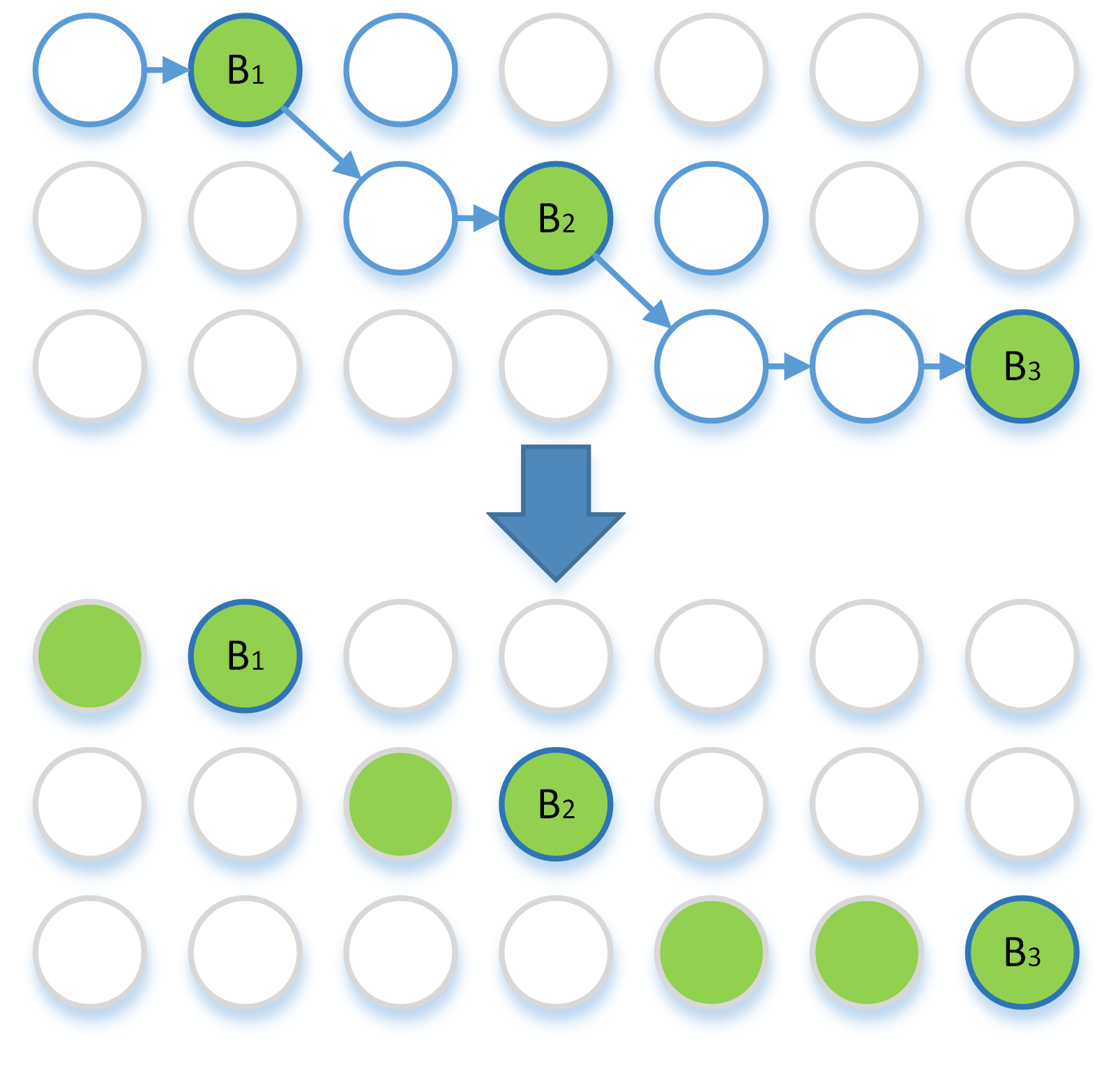}
  \caption{Illustration of the \AttName.}
  \label{fig:att}
\end{figure}

Based on the duration prediction of the input tokens, {\AttName} tries to align the text input and the mel spectrum frames, using the hidden states from text and mel encoders. As shown in Fig. \ref{fig:att}, {\TokenName} boundaries are monotonically searched in mel spectrum, then the alignments of frames between boundaries are filled by copying the alignments of the corresponding boundaries.

\subsection{Attention formulation} \label{subsec:\AttName}

In this section, we introduce our novel attention mechanism to monotonically search the {\TokenName} boundaries in the mel spectrum.
Let $\{x_i\}={x_1,x_2,...,x_{I}}$ be the {\TokenName} sequence with the length $I$, and $\{y_j\}={y_1,y_2,...,y_{J}}$ be the corresponding mel spectrum frames with the length $J$. Following the attention mechanism  \cite{vaswani2017attention}, scaled dot production is used to calculate the energy $e_{i,j}$ of each {\TokenName} and mel frame pair: 
\begin{equation}
  e_{i,j} = \exp(\frac{Q_i \cdot K_j}{\sqrt{\mathbf{d}_{QK}}})
  \label{eq:energy}
\end{equation}
where $\mathbf{d}_{QK}$ is the dimension of query $Q_i$ and key $K_j$. Here we take {\TokenName} $x_i$ as query and mel frame $y_j$ as key. Supposing that the max duration number of a {\TokenName} is $D$ (we use $D=20$ in our experiments), the boundary probability $\alpha_{i,j}$ that the mel frame $y_j$ is the boundary of the text token $x_i$ can be calculated as:
\begin{equation}
\begin{aligned}
    \alpha_{i,j}=&P(B_i=j)\\
    = & \sum_{k= max(j-D,0)}^{j-1}P(B_{i-1}=k) P(B_i=j|B_{i-1}=k)
  \label{eq:alpha}
\end{aligned}
\end{equation}
where $P(B_{i-1}=k)$ is the probability of that the {\TokenName} $x_{i-1}$ stops at frame $y_k$, and $P(B_i=j|B_{i-1}=k)$ is the conditional probability that $y_j$ is the boundary of $x_i$ given the boundary ($y_k$) of the previous {\TokenName} $x_{i-1}$. The conditional boundary probability $P(B_i=j|B_{i-1}=k)$ is calculated as:
\begin{equation}
\begin{aligned}
    P(B_i  =j|B_{i-1}=k)
    &=\frac{e_{i,j}}{\sum_{m=k+1}^{min(k+D,J)}e_{i,m}}.
\end{aligned}
  \label{}
\end{equation}
The probability of the special case that the boundary index is 0 is defined as:
\begin{equation}
    P(B_{i'}  = 0) = 
    \begin{cases}
        1 &\text{if ${i'}=0$ }\\
        0 &\text{otherwise}.
    \end{cases}
\end{equation}

Following the properties of monotonicity and continuity, the alignment probability $\beta_{i,j}$, indicating $y_j$ is aligned to $x_i$, is the probability that the boundary of $x_i$ is not before $y_j$, and the boundary of $x_{i-1}$ is before $y_j$:
\begin{equation}
\begin{aligned}
    \beta_{i,j}=&P(B_{i-1} < j \le B_i)\\
    = &\sum_{k=max(j-D,0)}^{j-1} P(B_{i-1}=k) P(B_i \ge j|B_{i-1}=k)
\end{aligned}
  \label{}
\end{equation}
where 
\begin{equation}
\begin{aligned}
     P(B_i \ge j|B_{i-1}=k) &= \sum_{j'=j}^{min(k+D, J)} P(B_i =j'|B_{i-1}=k)\\
\end{aligned}
  \label{}
\end{equation}

Based on the alignment probability  $\beta$,  we can tile the text hidden states to get the decoder input for the $j^{th}$ frame as:
\begin{equation}
\begin{aligned}
    \tilde{h_j}=\sum_{i=1}^{I}\beta_{i,j} h_i 
\end{aligned}
  \label{}
\end{equation}

\subsection{Encourage discreteness}

To extract the {\TokenName} boundaries in the mel spectrum, we want to concentrate the frames to only one {\TokenName}. Following Gumbel-Softmax \cite{maddison2016concrete,jang2016categorical}, two measures (the Gumbel noise $G_i$ and the temperature $\tau_i$) are introduced, and the energy $e_{i,j}$ in Eq. \ref{eq:energy} is rewritten as:
\begin{equation}
  e_{i,j}^{G} = \exp(\frac{Q_i \cdot K_j}{\sqrt{\mathbf{d}_{QK}}\tau_{i}}+\frac{G_{i,j}}{\tau_{i}})
  \label{eq:energy_g}
\end{equation}
The Gumbel noise $G_i$  is transformed by a random variable sampled from a uniform distribution:
\begin{equation}
\begin{aligned}
    U_{i,j} &\sim U(0,1) \\ 
    G_{i,j} &= -log(-log(U_{i,j}))\\
\end{aligned}
  \label{}
\end{equation}
The temperature $\tau_i$ is a non-negative scalar  to control the sharpness of the distribution: the closer $\tau_i$ is to 0, the sharper this distribution becomes . Instead of using $\tau_i$ anneals from $1$ to $0.1$, we use $\tau_{i} \sim U(0.1,\tau_{max})$, where $\tau_{max}$ linearly anneals from $1$ to $0.1$ in the training procedure, to make the discreteness process  smoother and more efficient.

\subsection{Frame interlacement for acceleration}
Since the adjacent mel frames are similar, to improve the training speed, we select the first one out of every 2 frames to calculate the energy $e$, the boundary probability $\alpha$ and the alignment probability $\beta$, which are then recovered to the original length by inserting 0 to the intervals for $\alpha$, and repeating each frame twice for $\beta$. 

\subsection{Inference} \label{subsec:infer}
Instead of sampling a random number for $\tau_i$, a fixed number $0.1$ is used during the inference, and the Gumbel noise is removed.
Besides, after each boundary probability $\alpha_{i,j}$ is calculated, an $onehot(\cdot)$ function is used to select the frame with the highest mass as the hard boundary for $x_i$:
\begin{equation}
\hat{\alpha}_{i,j}=onehot(\alpha_{i,j})
\end{equation}
then $\hat{\alpha}_{i}$ is used to calculate the boundary probabilities $\alpha_{i+1}$ for the next text token.

To ensure $\sum_{i=1}^{I} d_i=J$, where $d_i$ is the duration of $x_i$, we set the duration of the last text token as $d_{I}=J-\sum_{i=1}^{I-1} d_i$. Since there is another constraint that the {\TokenName} durations should be smaller than the maximum duration $D$, the sample will be abandoned if $d_{I}>D$ during inference, and we use all the success samples for the non-autoregressive TTS model training. Such a process can filter bad samples with noise from the training data.

\section{Experiment}

In this section,  we train our proposed new aligner {\ModelName} and extract {\TokenName} durations, based on which a {\NarName} \cite{ren2019fastspeech} is trained to synthesize the mel spectrum and generate the audio. Mean option score (MOS) is leveraged for the evaluation. 

\subsection{Dataset}
Our experiments are conducted on LJSpeech dataset \cite{ljspeech17}, which includes 13,100 clips (splitted into 12,600/250/250 parts for train/dev/eval respectively) and totally 24 hours. 

\subsection{Setup}
Our {\ModelName} is implemented based on ESPnet \cite{watanabe2018espnet,hayashi2019espnettts}, an open-sourced repository available from GitHub, in which pretrained vocoders are also available. We use \emph{ljspeech.wavenet.mol.v1}, a pretrained Wavenet,  as our vocoder. 
We also implement a "DistributedDataParallel" model in pytorch  to improve the training speed.
The frame size of each batch is 15000 (80 samples per batch on average). Adam optimizer is employed for model training, with the same setting and the Noam decay method \cite{vaswani2017attention}. Our model is trained for 400 epoches, and the total training time is 14 hours ($\sim$1 iter/sec) with 4 Tesla P40 GPUs (24G RAM), which reduces 30\% training time comparing to TransformerTTS (20 hours).

\subsection{Visualization of aligning results}
\begin{figure}[tb]
  \centering
    \includegraphics[width=0.99\linewidth]{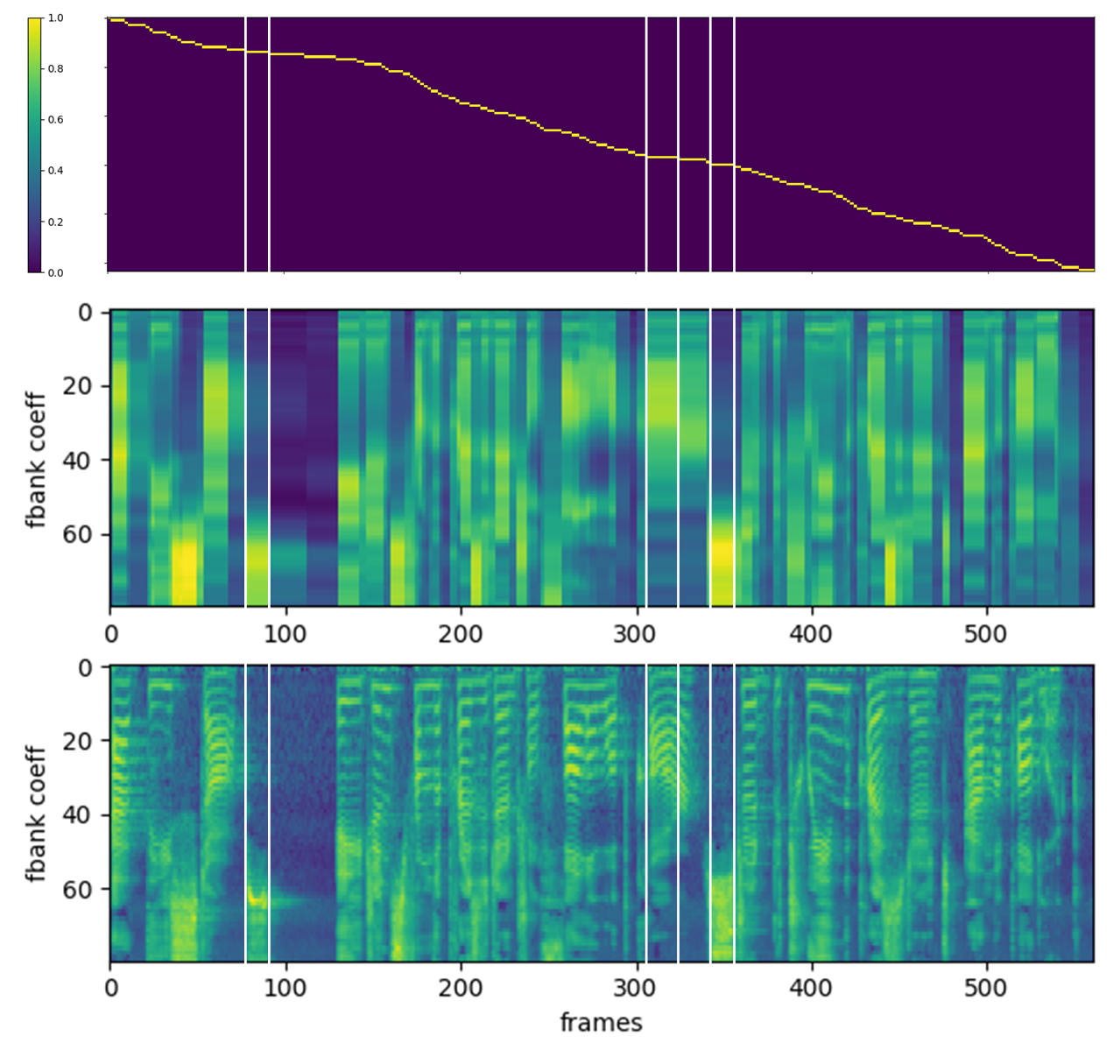}
  \caption{Visualization of aligning results. The top figure is the alignment matrix, the middle one is the reconstructed mel by the linear layer in {\ModelName}, and the bottom one is the ground truth mel spectrum. }
  \label{fig:alignres}
\end{figure}

As shown in Fig. \ref{fig:alignres}, the aligning results can be visualized by the alignment matrix and the reconstructed mel spectrum generated by the linear projection layer in {\ModelName}. From the alignment matrix, we can find that the two properties of monotonicity and continuity are well obeyed: each frame focuses on one {\TokenName}, and all {\TokenName}s are covered by continuous frames.
As for reconstructed mel spectrum, although the reconstructed mel spectrum is blurred, the {\TokenName} boundaries are still clear and  accurate, from which it can be derived that {\ModelName} has a tremendous ability in locating the token boundary.

\subsection{TTS model and baselines}
With the extracted {\TokenName} durations by {\ModelName}, a {\NarName} is trained to evaluate the duration quality. 
{This model is compared with another {\NarName}, of which the duration is provided by TransformerTTS. 
Therefore, our experiments, as shown in Table \ref{tab:mos}, consist of following models:
1) "GT Mel Wavs": the synthesized waveforms with the ground truth mel spectrums, aiming to exclude the effect of the vocoder; 2) "TransformerTTS": the autoregressive model; 3) "TF dur + {\NarName}" and "MB dur + {\NarName}": two non-autoregressive TTS models both using {\NarName} as the TTS model, while "TF dur + {\NarName}" uses the duration from TransformerTTS, and  "MB dur + {\NarName}" uses that of our proposed MoBoAligner. In these two models, we directly use the recipe in Espnet, which doesn't leverage the sequence-level knowledge distillation, and is different from the original Fastspeech in \cite{ren2019fastspeech}.
}
Both {\NarName} and TransformerTTS are of the same attention size (384), attention head number (2), and FFN hidden size (1536). 

\subsection{Result}
The results of the mean option score (MOS) are shown in  Table \ref{tab:mos}.
The result with ground truth mel spectrum (GT Mel Wavs) achieves close quality with recordings followed by the autoregressive model TransformerTTS.
Two non-autoregressive models are worse than TransformerTTS since the synthesized audios sound hoarse sometimes.  Leveraging the same {\NarName} models, replacing the TransformerTTS aligner with our proposed \ModelName can still get a significant improvement ({\MbMos} vs {\FsMos}) on the MOS score. 

\begin{table}[!t]
\begin{center}
\begin{tabular}{c|c|c}
    \toprule
    \textbf{Model}  &   \textbf{MOS}    &   CI\\
    \midrule
    Recording   &   {\RcMos}    &   0.10\\
    GT Mel Wavs &   {\GtMos}    &   0.07\\
    \midrule
    TransformerTTS  &   {\TfMos}    &   0.09\\
    \midrule
    TF dur + {\NarName} (w/o distillation) &   {\FsMos}    & 0.10\\
    MB dur + {\NarName} (w/o distillation) &   {\MbMos}    & 0.10\\
    \bottomrule
\end{tabular}
\end{center}
\caption{MOS test results.   "CI" means the confidence interval radius with confidence level 0.95.}
\label{tab:mos}
\end{table}

\subsection{Comparison with \FA}
To extract the text token duration, there is another way called {\FA}. We also employ an internal HMM-based {\FA} model, trained with the speech recognition data of $\sim$10,000 hours, and adapted with LJSpeech dataset. Although this {\FA} tool uses larger training data and more complex training procedures, the final result using {\ModelName} achieves the same MOS with that using this {\FA} tool.

\section{Conclusions}
In this paper, we propose {\ModelName}, which employs a novel attention mechanism to monotonically search the {\TokenName} boundaries in the mel spectrum, and extract durations of high accuracy. 
Our model is trained under the instruction of MSE loss with an end-to-end framework. The ability in locating token boundaries can be proven in two aspects. On the one hand, the reconstructed mel spectrum has clear token boundaries among frames, although the detail is blurred.
On the other hand, our model provides better durations which is revealed by the improvement of the non-autoregressive TTS model on MOS ({our {\MbMos} comparing to baseline's {\FsMos}}). Besides, our model is task-specified and lightweight, which reduces the parameter number by 45\% and the training time consuming by 30\%.

% ...

\bibliographystyle{IEEEtran}

\bibliography{mybib}


\end{document}